\documentclass[12pt]{article}

\usepackage{amssymb}
\usepackage{amsmath}
\usepackage{amsfonts}
\usepackage{amsbsy}

\begin{document}

\begin{titlepage}

\begin{center}

\vspace*{-10ex}
\hspace*{\fill} KAIST-TH/2005-01

\vskip 1.5cm

\Huge{From the Spectrum to Inflation :\\
An Inverse Formula for the\\ General Slow-Roll Spectrum}

\vskip 1cm

\large{ Minu Joy$^1$\footnote{minujoy@muon.kaist.ac.kr} \hspace{0.2cm} Ewan D.
Stewart$^{1,2}$\footnote{On sabbatical leave from Department of Physics, KAIST,
Daejeon, Republic of Korea} \hspace{0.2cm} Jinn-Ouk Gong$^1$ \hspace{0.2cm}
Hyun-Chul Lee$^1$
\\ \vspace{0.5cm}
{\em ${}^1$ Department of Physics, KAIST, Daejeon, Republic of Korea
\\ \vspace{0.2cm}
${}^2$ Department of Physics and Astronomy, University of Canterbury,\\
Christchurch, New Zealand} }

\vskip 0.5cm

\today

\vskip 1.2cm

\end{center}

\begin{abstract}

We propose a general inverse formula for extracting inflationary parameters from
the power spectrum of cosmological perturbations. Under the general slow-roll
scheme, which helps to probe the properties of inflation in a model independent
way, we invert the leading order, single field, power spectrum formula. We also
give some physically interesting examples to demonstrate its wide applicability
and illuminate its properties.

\end{abstract}

\end{titlepage}

\setcounter{page}{0}

\newpage

\setcounter{page}{1}

\section{Introduction}

It is believed that the power spectra of both the cosmic microwave background
radiation and the large scale structure of the observable universe have evolved
from a common primordial spectrum of curvature perturbations produced during
inflation \cite{ps}. Through the observation of these spectra we can study the
properties of inflation. For example, results from the Wilkinson Microwave
Anisotropy Probe \cite{wmap} and Sloan Digital Sky Survey \cite{sdss} provide a
remarkable set of data, which places strong restrictions on models of inflation
\cite{wmapsdss}.

There have been many studies on the accurate evaluation of the power spectrum,
provided that the inflationary parameters are given. But practically we observe
the power spectrum first, and then seek an appropriate inflationary model from
which the observed spectrum could result \cite{recon}. In the context of the
standard slow-roll approximation, such inversion is straightforward
\cite{recon}. However, the standard slow-roll approximation makes strong
assumptions about the properties of inflation, which have not yet been
confirmed observationally. Thus we are lacking a model independent way of
extracting useful information about inflationary parameters from the power
spectrum.

The general slow-roll approximation \cite{gsr} was introduced to eliminate
these unjustified assumptions of standard slow-roll, instead essentially only
relying on the observed approximate scale invariance of the spectrum.
See Refs.~\cite{SalmanUniform,SchwarzWKB} for some other alternatives to the
standard slow-roll approximation.
The one limitation of general slow-roll is that it does not cover cases where
super-horizon effects are dominant \cite{misao}.
Also, in this paper, we limit ourselves to the case of single field models of
inflation, though the single field general slow-roll formulae can also be
applied to some multi-field models of inflation \cite{kenji}.
These extensions will be treated thoroughly in a separate publication
\cite{yokoyama}.

In this paper, we invert the leading order single field general slow-roll
formula for the power spectrum \cite{gsr} to obtain a formula, Eq.~(\ref{inv}),
for inflationary parameters in terms of the primordial power spectrum.
See Ref.~\cite{SalmanInverse} for an alternative method of inversion that also
does not rely on the standard slow-roll approximation.
In Section~\ref{inverseformulae} we give our inverse, in Section~\ref{examples}
we give some examples which illustrate the use and properties of our inverse,
and in the Appendix we give some alternative forms for some of our formulae.

\section{General slow-roll formulae}
\label{inverseformulae}

For single field inflationary models, it is convenient to express inflationary
quantities in terms of
\begin{equation}
\label{f}
f = \frac{2\pi a \xi \dot\phi}{H}
\, ,
\end{equation}
where $\xi = - \int \frac{dt}{a} = \frac{1}{aH} \left( 1 - \frac{\dot{H}}{H^2} +
\cdots \right)$ is minus the conformal time \cite{gsr,jo1}. We think of $f$ as a
function of $\ln\xi$ so that $f' \equiv df/d\ln\xi$.

To leading order in the general slow-roll approximation, the spectrum can be
expressed as \cite{gsr}
\begin{equation}
\label{PW}
\ln\mathcal{P}(k) = \int_0^\infty \frac{d\xi}{\xi} \left[ -k\xi \, W'(k\xi)
\right] \left[ \ln \left( \frac{1}{f^2} \right) + \frac{2}{3} \frac{f'}{f}
\right] \, .
\end{equation}
There are a variety of other forms for this formula, some of which are given in
the Appendix. The window function $- x \, W'(x)$ is given by \footnote{Note
that we define $W(x)$ with an extra $-1$ compared with our previous works
\cite{gsr}.}
\begin{equation}
W(x) = \frac{3\sin(2x)}{2x^3} - \frac{3\cos(2x)}{x^2} - \frac{3\sin(2x)}{2x} -
1 \, .
\end{equation}
It has the asymptotic behavior
\begin{equation}
\lim_{x \rightarrow 0} W(x) = \frac{2}{5} x^2 + \mathcal{O}(x^4)
\, ,
\end{equation}
the window property
\begin{equation}
\int_0^\infty \frac{dx}{x} \left[-x\,W'(x)\right] = 1
\, ,
\end{equation}
and the degeneracy
\begin{equation}
\label{Wdeg}
\int_0^\infty \frac{dx}{x} \left[-x\,W'(x)\right] \frac{1}{x} = 0
\, .
\end{equation}

\subsection{Inverse formula}

Our inverse formula is
\begin{equation}
\label{inv}
\ln \left( \frac{1}{f^2} \right) = \int_0^\infty \frac{dk}{k} \, m(k\xi)
\ln\mathcal{P} \, ,
\end{equation}
where
\begin{equation}
m(x) = \frac{2}{\pi} \left[ \frac{1}{x} - \frac{\cos(2x)}{x} - \sin(2x) \right]
\, .
\end{equation}
It has the asymptotic behavior
\begin{equation}
\lim_{x \rightarrow 0} m(x) = \frac{4}{3\pi} x^3 + \mathcal{O}(x^5)
\end{equation}
and the window property
\begin{equation}
\label{mwin}
\int_0^\infty \frac{dx}{x} \, m(x) = 1
\, .
\end{equation}
It is straightforward to derive Eq.~(\ref{inv}) from Eq.~(\ref{PW}) using the
key identity
\begin{equation}
\label{id}
\int_0^\infty \frac{dk}{k} \, m(k\zeta) \, W(k\xi) = \left(
\frac{\zeta^3}{\xi^3} - 1 \right) \theta(\xi-\zeta) \, ,
\end{equation}
where $\theta(x)=0$ for $x<0$ and $\theta(x)=1$ for $x>0$.
An alternative form of the inverse formula is given in the Appendix.

\section{Examples}
\label{examples}

\subsection{Standard slow-roll approximation}

In the context of standard slow-roll, the power spectrum has the form
\begin{equation}
\ln\mathcal{P} = \ln\mathcal{P}_\diamond + (n_\diamond - 1) \ln \left(
\frac{k}{k_\diamond} \right) + \cdots \, ,
\end{equation}
where $k_\diamond$ is some reference wavenumber. Applying our inverse formula,
Eq.~(\ref{inv}), gives
\begin{equation}
\label{ssrinv}
\ln \left( \frac{1}{f^2} \right) = \ln\mathcal{P}_\diamond + \alpha (n_\diamond
- 1) - (n_\diamond - 1) \ln(k_\diamond\xi) + \cdots \, ,
\end{equation}
where we have used Eq.~(\ref{mwin}) and
\begin{equation}
\int_0^\infty \frac{dx}{x} \, m(x) \ln x = \alpha
\, ,
\end{equation}
where $\alpha = 2 - \ln 2 - \gamma \simeq 0.7296$. We see that
Eq.~(\ref{ssrinv}) reproduces the standard slow-roll inverse, which is
trivially obtained from the standard slow-roll formulae \cite{jo1}
\begin{equation}
\ln\mathcal{P} = \ln \left( \frac{1}{f_\star^2} \right) - 2 \alpha
\frac{f_\star'}{f_\star} + 2 \frac{f_\star'}{f_\star} \ln(k\xi_\star) + \cdots
\end{equation}
and
\begin{equation}
n - 1 = 2 \frac{f_\star'}{f_\star} + \cdots
\, ,
\end{equation}
where $\xi_\star$ is an arbitrary evaluation point, usually taken to be around
horizon crossing.

\subsection{Power law}

Consider a spectrum of the form
\begin{equation}
\label{powerP}
\ln \mathcal{P} = \ln \mathcal{P}_0 - A k^\nu
\end{equation}
with $\nu > 0$. In the limit $\nu \ll 1$, this reproduces a significant subset
of standard slow-roll spectra. However, we have in mind that instead it is $A$
that is sufficiently small to give the required approximate scale invariance
over observable scales. Besides being motivated by simplicity, such a form for
the power spectrum is well motivated by inflationary model building
\cite{kenji}.

Substituting into our inverse formula, Eq.~(\ref{inv}), we obtain
\begin{equation}
\label{powerfP}
\ln \left( \frac{1}{f^2} \right) = \ln \mathcal{P}_0 - A \, C(\nu) \,
\xi^{-\nu} \, ,
\end{equation}
where
\begin{eqnarray}
C(\nu) & = & \int_0^\infty \frac{dx}{x} \, m(x) \, x^\nu
\nonumber \\
& = & \frac{2^{1-\nu}}{\pi} \frac{\Gamma(2+\nu)}{\nu(1-\nu)} \sin \left(
\frac{\pi\nu}{2} \right) \ \ \ \mbox{for} \ \ \nu < 1 \, .
\end{eqnarray}
For $\nu \geq 1$, the inverse diverges. This is to be expected as
Eq.~(\ref{Wdeg}) shows that the forward formula, Eq.~(\ref{PW}), is degenerate
for $\nu = 1$.

For a better understanding of this degeneracy, we can start from an inflationary
scenario \cite{kenji} which leads to a power spectrum of the form of
Eq.~(\ref{powerP}). In such a scenario, $f$ is calculated to be
\begin{equation}
\label{powerf}
\ln \left( \frac{1}{f^2} \right) = \ln \left( \frac{1}{f_\infty^2} \right) - B
\xi^{-\nu}
\end{equation}
with $\nu > 0$. Then, calculating the power spectrum using Eq.~(\ref{Pw}), we
obtain
\begin{equation}
\label{powerPf}
\ln \mathcal{P} = \ln \left( \frac{1}{f_\infty^2} \right) - B \, D(\nu) \,
k^\nu \, ,
\end{equation}
where
\begin{eqnarray}
D(\nu) & = & \int_0^\infty \frac{dx}{x} \left[ - x \, w'(x) \right] x^{-\nu}
\nonumber \\
& = & 2^\nu \frac{\Gamma(2-\nu)}{1+\nu} \cos \left( \frac{\pi\nu}{2} \right) \
\ \ \mbox{for} \ \ \nu < 2 \, .
\end{eqnarray}
For $\nu < 1$ everything is straightforward and similar to the standard
slow-roll case $\nu \ll 1$. For $\nu = 1$ we have $D(1) = 0$ so that the first
order general slow-roll formula becomes degenerate. Thus we cannot expect an
inverse to exist for $\nu \geq 1$. For $1 < \nu < 2$, the super-horizon tail of
the window function $- x \, w'(x)$ dominates, but the general slow-roll
formula, Eq.~(\ref{Pw}), still converges. For $\nu \geq 2$, the general
slow-roll formula diverges indicating the need for super-horizon methods
\cite{misao,yokoyama}.

Finally, from Eq. (\ref{powerfP}) or~(\ref{powerPf}), we have
\begin{equation}
\ln\mathcal{P}_0 = \ln \left( \frac{1}{f_\infty^2} \right) \, ,
\end{equation}
and noting that
\begin{equation}
C(\nu) \, D(\nu) = 1
\, ,
\end{equation}
we see the consistency of Eqs.~(\ref{powerP}) and~(\ref{powerfP}) with
(\ref{powerf}) and~(\ref{powerPf}).

\subsection{Linear potential with a sharp step}

As our final example, we consider a linear potential of slope $A$ with a sharp
downward step at $\phi_\mathrm{s}$. See Ref.~\cite{step} for a more complete
discussion of this example. The potential is
\begin{equation}
V(\phi) = V_0 [ 1 - A (\phi - \phi_\mathrm{s}) - a \, \theta(\phi - \phi_\mathrm{s}) ]
\, ,
\end{equation}
where $A$ determines the slope of the potential and $a$ the height of the step.
The potential is not smooth and we have a singularity at the position of the
step. Therefore we cannot apply the standard slow-roll approximation since
$\dot\phi$ is changing drastically at the step. But we can apply the general
slow-roll approximation to calculate the power spectrum. We assume $A \ll 1$ so
that de Sitter space is a good approximation, and $a V_0 \ll A^2$ so that the
step only generates a small departure from scale invariance. Then, we have
\begin{equation}
\label{stepf}
\ln \left( \frac{1}{f^2} \right) = \ln \left( \frac{V_0}{12\pi^2A^2} \right) -
2\frac{aV_0}{A^2} \left( \frac{\xi}{\xi_\mathrm{s}} \right)^3
\theta(\xi_\mathrm{s} - \xi) \, .
\end{equation}
Using Eq.~(\ref{PW}), the power spectrum is
\begin{equation}
\ln \mathcal{P} = \ln \left( \frac{V_0}{12\pi^2A^2} \right) + \frac{2}{3}
\frac{aV_0}{A^2} k\xi_\mathrm{s} \, W'(k\xi_\mathrm{s}) \, .
\end{equation}
Now, using our inverse, Eq.~(\ref{inv}), to invert this power spectrum we
recover Eq.~(\ref{stepf}) by using Eq.~(\ref{id}).
This again illustrates the power of the general slow-roll approach compared
with the standard slow-roll approach.

\subsection*{Acknowledgements}

We thank Jai-chan Hwang, Kenji Kadota, Misao Sasaki and Takahiro Tanaka for
helpful discussions, and the Yukawa Institute for Theoretical Physics for
hospitality while this work was in progress. JG is grateful to Jai-chan Hwang
and Misao Sasaki for encouragement. This work was supported in part by ARCSEC
funded by the Korea Science and Engineering Foundation and the Korean Ministry
of Science, the Korea Research Foundation grant KRF PBRG 2002-070-C00022, Brain
Korea 21, and an Erskine Fellowship of the University of Canterbury.

\section*{Appendix}

In this appendix we give some alternative formulae for the spectrum and the
inverse.

If $f_\infty \equiv \lim_{\xi \rightarrow \infty} f$ exists, as for example in
the models of Ref.~\cite{kenji}, then we have the following form
\begin{equation}
\label{Pw}
\ln\mathcal{P} = \ln \left( \frac{1}{f_\infty^2} \right) + \int_0^\infty
\frac{d\xi}{\xi} \left[ - k\xi \, w'(k\xi) \right] \ln \left(
\frac{f_\infty^2}{f^2} \right) \, ,
\end{equation}
where \footnote{Note that we define $w(x)$ with an extra $-1$ compared with our
previous works \cite{gsr}.}
\begin{equation}
w(x) = \frac{\sin(2x)}{x} - \cos(2x) - 1
\, .
\end{equation}
It has the asymptotic behavior
\begin{equation}
\lim_{x \rightarrow 0} w(x) = \frac{2}{3} x^2 + \mathcal{O}(x^4)
\, ,
\end{equation}
the almost window property
\begin{equation}
\int_0^\infty \frac{dx}{x} \left[-x\,w'(x)\right] = 1 + \cos(2\infty)
\, ,
\end{equation}
and the degeneracy
\begin{equation}
\label{wdeg}
\int_0^\infty \frac{dx}{x} \left[-x\,w'(x)\right] \frac{1}{x} = 0
\, .
\end{equation}
It is related to $W(x)$ by
\begin{equation}
w(x) = W(x) + \frac{x}{3} \, W'(x)
\, ,
\end{equation}
and Eq.~(\ref{id}) gives
\begin{equation}
\int_0^\infty \frac{dk}{k} \, m(k\zeta) \, w(k\xi) = - \theta(\xi-\zeta)
\, .
\end{equation}

An alternative form for the spectrum with a particularly simple window function
is
\begin{equation}
\ln\mathcal{P} = \int_0^\infty \frac{d\xi}{\xi} \left[ -k\xi \, v'(k\xi)
\right] \left[ \ln \left( \frac{1}{f^2} \right) - 2 \frac{f'}{f} \right] \, ,
\end{equation}
where
\begin{equation}
v(x) = \frac{\sin(2x)}{2x} - 1
\, .
\end{equation}
It has the asymptotic behavior
\begin{equation}
\lim_{x \rightarrow 0} v(x) = - \frac{2}{3} x^2 + \mathcal{O}(x^4) \, ,
\end{equation}
window property
\begin{equation}
\int_0^\infty \frac{dx}{x} \left[-x\,v'(x)\right] = 1
\, ,
\end{equation}
and is related to $w(x)$ by
\begin{equation}
w(x) = v(x) - x \, v'(x)
\, .
\end{equation}

An alternative form for the inverse with a particularly simple window function
is
\begin{equation}
\ln \left( \frac{1}{f^2} \right) = \int_0^\infty \frac{dk}{k} \, n(k\xi) \left[
\ln\mathcal{P} + \frac{d\ln\mathcal{P}}{d\ln k} \right] \, ,
\end{equation}
where
\begin{equation}
n(x) = \frac{1}{\pi} \left[ \frac{1}{x} - \frac{\cos(2x)}{x} \right]
\, .
\end{equation}
It has the asymptotic behavior
\begin{equation}
\lim_{x \rightarrow 0} n(x) = - \frac{2}{\pi} x + \mathcal{O}(x^3)
\, ,
\end{equation}
window property
\begin{equation}
\int_0^\infty \frac{dx}{x} \, n(x) = 1
\, ,
\end{equation}
and is related to $m(x)$ by
\begin{equation}
m(x) = n(x) - x \, n'(x)
\, .
\end{equation}

\end{document}